\documentclass[3p,times,twocolumn]{elsarticle}
\usepackage{slashed}
\usepackage{amsmath}

\makeatletter
\DeclareRobustCommand{\cev}[1]{%
  \mathpalette\do@cev{#1}%
}
\newcommand{\do@cev}[2]{%
  \fix@cev{#1}{+}%
  \reflectbox{$\m@th#1\vec{\reflectbox{$\fix@cev{#1}{-}\m@th#1#2\fix@cev{#1}{+}$}}$}%
  \fix@cev{#1}{-}%
}
\newcommand{\fix@cev}[2]{%
  \ifx#1\displaystyle
    \mkern#23mu
  \else
    \ifx#1\textstyle
      \mkern#23mu
    \else
      \ifx#1\scriptstyle
        \mkern#22mu
      \else
        \mkern#22mu
      \fi
    \fi
  \fi
}

\makeatother

\usepackage{ecrc}

\volume{00}

\firstpage{1}

\journalname{Nuclear and Particle Physics Proceedings}

\runauth{}

\jid{nppp}

\jnltitlelogo{Nuclear and Particle Physics Proceedings}

\usepackage{amssymb}

\usepackage[figuresright]{rotating}

\begin{document}

\begin{frontmatter}



\dochead{}

\title{The initial state and hard probes: a brief review}


\author{Raju Venugopalan}

\address{Physics Department, Brookhaven National Laboratory, Bldg.\ 510A, Upton, NY 11973, USA}

\begin{abstract}
We provide a brief review of some of the recent developments in our understanding of the initial state in ultra-relativistic heavy-ion collisions. 

\end{abstract}


\end{frontmatter}


\section{Introduction}
\label{sec:intro}

In this talk, we will provide a brief, and therefore of necessity, incomplete review of recent developments in our understanding of the initial state in heavy-ion collisions. Acknowledging the scope of the conference, I will address a few examples where hard probes can provide insight into the nature of the initial state. 

In studying the real time dynamics of strongly interacting gauge theories, there are two clean asymptotic limits where one can obtain clear answers to well posed questions. One is in the limit of large t'Hooft coupling $g^2 N_c$ and large $N_c$, where a duality may be established between correlation functions in strongly coupled  supersymmetric Yang-Mills theories with $N=4$ supercharges and weakly coupled gravity in a 10 dimensional $AdS5\times S5$ spacetime~\cite{Maldacena:1997re}. The other clean limit is that of very weak coupling $g\rightarrow 0$ in QCD but $g^2 f\sim 1$ where $f$ denotes the occupancy of gluon modes\footnote{An equivalent gauge invariant measure is the field strength squared in units of a hard scale of interest.}~\cite{Gribov:1984tu,Mueller:1985wy,McLerran:1993ni,McLerran:1993ka,McLerran:1994vd}. Our focus here will be on the strongly correlated gluodynamics of the initial state and early time dynamics in weak coupling.

\section{The CGC wavefunction}
The strongly correlated dynamics of saturated gluons in hadron wavefunctions is described by the CGC effective theory~\cite{Gelis:2010nm}. The CGC initial state is a highly Lorentz contracted gluon shock wave that is ``lumpy" in the transverse plane on a scale $1/Q_S$. This saturation scale is the color screening length measured by a quark-antiquark dipole probe. $Q_S$ grows with energy (or decreasing Bjorken $x$) and with nuclear size; its rate of growth is described by the Balitsky-Kovchegov (BK) renormalization group (RG) equation~\cite{Balitsky:1995ub,Kovchegov:1999yj}. When $Q_S \gg \Lambda_{\rm QCD}$, the many-body parton dynamics inside nuclear wavefunctions can be described in weak coupling. Because the dynamics is captured by one emergent scale, the CGC framework has enormous predictive power in a regime of QCD where the equations describing color fields are strongly non-linear. 

In high energy scattering, cross-sections are described in terms of dipole, quadrupole and in principle multipole products of lightlike Wilson line correlators. These appear in both Deeply Inelastic Scattering (DIS) and hadron-hadron scattering and their evolution with energy is described by the B-JIMWLK hierarchy of equations~\cite{Balitsky:1995ub,JalilianMarian:1997gr,Iancu:2000hn}; the BK equation is a closed form simplification of the equation for dipole correlators in the large $N_c$ and large $A$ limit. There has been significant progress in the last few years in extending the B-JIMWLK hierarchy to NLO; progress in this direction is reviewed in the plenary lecture by Beuf~\cite{Beuf}. 

For practical applications, there are two widely adopted approaches to describe gluon saturation. One is within the framework of the IP-Sat model~\cite{Kowalski:2003hm}. The key ingredient in this model is the previously noted dipole cross-section. Its behavior depends on $Q_S (x,b)$, which is now a function of $x$ and impact parameter $b$. The latter dependence accounts for the fact that color screening may vary depending on whether the center or periphery of the proton is probed by the quark-antiquark dipole. In the IP-Sat model, $Q_S(x,b$) is determined from fits to HERA inclusive and exclusive data~\cite{Rezaeian:2012ji}. A recent development in this regard is the neat observation by Mantysaari and Schenke~\cite{Mantysaari:2016ykx} that a spherical impact parameter profile of glue in the proton does not describe HERA data on incoherent exclusive $J/\Psi$ production. This measurement, which is sensitive to 
fluctuations of the dipole cross-section, is better fit by non-spherical profiles, such as for instance those that might be generated by bremsstrahlung of gluons from constituent quarks~\cite{Schlichting:2014ipa}. 

The other phenomenological approach is one where the dipole cross-section is determined within the framework of the BK equation. The state of the art here is the NLO BK equation; its numerical implementation is discussed in the talk by Lappi~\cite{Lappi}. While this approach is better motivated from first principles, including impact parameter dependence in a reliable manner remains challenging and complicates phenomenological analyses. 

\section{Hadron-hadron collisions in the CGC framework}
Collisions at high energies, being those of lumpy gluon shocks, are not classified by the atomic number of the projectile $A$ or target $B$ but instead by the respective saturation scales and the typical transverse momenta involved~\cite{Gelis:2007kn}. Dilute-dilute collisions, defined as $Q_{S,A}^2/k_{T,A}^2 <<1$ and $Q_{S,B}^2/k_{T,A}^2 <<1$ can correspond to high transverse  momentum processes in nucleus-nucleus collisions or alternately, dynamics at moderate $k_T$ in proton-proton collisions. In this regime, if $x<<1$, the CGC matches smoothly to pQCD computations of hard processes; its definition as an effective field theory depends on it! Dilute-dense collisions correspond to 
$Q_{S,A}^2/k_{T,A}^2 <<1$ and $Q_{S,B}^2/k_{T,B}^2\sim 1$. In these kinematics, which corresponds for instance to final states in proton-nucleus (p+A) collisions or forward proton-proton (p+p) collisions, a hybrid pQCD/CGC description is feasible. Dynamics from the proton side is treated using collinear or $k_T$ factorization while that from the nuclear side includes high twist effects represented by Wilson line correlators. These last can be computed, as noted, using the BK/B-JIMWLK RG equations. Finally, dense-dense power counting corresponds to $Q_{S,A}^2/k_{T,A}^2\sim 1$, $Q_{S,B}^2/k_{T,B}^2\sim 1$. In this case, which is relevant for the bulk properties of a heavy-ion (A+A) collision, there is no small parameter.  However, the classical Yang-Mills equations describing the dynamics can be solved numerically in 2+1-D and 3+1-D, with the leading quantum fluctuations resummed into stochastic initial conditions~\cite{Gelis:2007kn,Dusling:2011rz,Epelbaum:2013waa}. 

\subsection{Dilute-dense results for p+A collisions}
A significant development is the treatment of the single inclusive hadron spectrum in p+A collisions beyond LO~\cite{Stasto:2016wrf}. The first NLO computations gave NLO results that improved agreement with data at low $p_T$ but gave unphysical results at $p_T$'s greater than a few GeV~\cite{Chirilli:2012jd,Stasto:2014sea,Stasto:2014sea}. This is because kinematical constraints become increasingly important in matching to collinear factorization at high $p_T$~\cite{Altinoluk:2014eka,Iancu:2016vyg}. As discussed in the talk by Yan Zhu~\cite{YanZhu}, the problem may be resolved by proper treatment of rapidity factorization schemes. 

Onium production in p+p and p+A collisions is successfully described in a CGC+NRQCD framework~\cite{Kang:2013hta}. For forward p+p and p+A, the dilute-dense framework, employing the running coupling BK equation, gives a good description of RHIC and LHC data at low $p_T$~\cite{Ma:2014mri,Ma:2015sia}. This framework smoothly matches to an NLO pQCD+NRQCD framework at higher $p_T$~\cite{Ma:2010jj}. An interesting conclusion of this study is that color octet channels dominate the $J/\Psi$ cross-section, with the color singlet contribution providing only a 10\% contribution in p+p collisions and at most 15-20\% of the cross-section in p+A collisions. Ducloe in his talk~\cite{Ducloe} showed that previous disagreement of data with CGC predictions~\cite{Fujii:2013gxa} arose from an improper treatment of the p+A geometry~\cite{Ducloue:2015gfa,Ducloue:2016pqr}. 

Both the Color Evaporation model (CEM) and NRQCD describe the p+A $J/\Psi$ data within uncertainties; since octet mechanisms dominate both descriptions, this is perhaps not too surprising. As discussed in several talks, in particular the plenary lecture by Ferreiro~\cite{Ferreiro}, the $\frac{\Psi^\prime}{J/\Psi}$ ratio can be described by rescattering; a very slight modification of the CEM model to account for initial state soft gluon comover exchanges describes the systematics of the data~\cite{Ma:2016exq,MVWZ}.

Benic in his talk~\cite{Benic,Benic:2016yqt} noted that a framework identical to the one for heavy quark pair production~\cite{Benic:2016uku} gives the leading contribution to photon production in p+A collisions. Specifically, Low's theorem shows that the cross-section factorizes into the cross-section for quark-antiquark pair production times a kinematical factor corresponding to photon bremsstrahlung for soft photons. Another interesting result is that at high $p_T$ this cross-section smoothly goes over into the collinearly factorized expression proportional to the nuclear gluon distribution. Thus not only can the nuclear gluon distribution be extracted in this process but higher twist contributions to the same can be quantified as well. This matching of the two frameworks suggests that concepts such as shadowing and energy loss are not mutually exclusive but can be viewed as leading and sub-leading contributions respectively in a systematic power counting scheme.

\subsection{The Glasma:recent developments}
The Glasma is the non-equilibrium QGP arising from the large occupancy of gluons in the initial state of dense-dense collisions~\cite{Kovner:1995ja,Kharzeev:2001ev,Lappi:2006fp}. Understanding its strongly correlated dynamics is key to understanding how the QGP is formed, as well as what the smallest systems are to which the concept of a thermalized QGP can be applied. 

Long-range rapidity correlations are sensitive to the early time dynamics of the Glasma~\cite{Dumitru:2008wn}. Two particle rapidity correlations computed in a CGC ``JIMWLK factorization" framework~\cite{Dusling:2009ni} show a predicted decorrelation whose magnitude is consistent with CMS data~\cite{Schenke:2016ksl}. Data over a wider rapidity range can help distinguish between differing models of long range rapidity correlations~\cite{Pang:2015zrq,Broniowski:2015oif}. 

How the Glasma thermalizes has long been an outstanding problem. In weak coupling, the initial classical configurations are boost invariant at $\tau=0^+$ but are unstable to quantum fluctuations. Very early time dynamics at $\tau < 1/Q_S$ were discussed in the talks of Fries and McDonald~\cite{Fries,McDonald}. Quantum fluctuations grow exponentially, and on a very short time scale of $\tau \sim Q_S^{-1}\ln(1/\alpha_S^2)$ are of the same magnitude of the classical fields~\cite{Romatschke:2005pm,Romatschke:2006nk}. Depending on differing initial conditions, the gauge fields at this time can be prolate or oblate to differing degree in the $p_T-p_Z$ momentum plane~\cite{Berges:2014yta}. The subsequent evolution of the Glasma, described by solutions of classical-statistical classical Yang-Mills equations, is a competition between the rapid expansion which squeezes distributions towards the infrared in $p_Z$, and scattering between gluons, which attempts to broaden the $p_Z$ distribution. 

Real time numerical simulations for an $SU(2)$ gauge theory on $256^2 \times 4096$ lattices reveal~\cite{Berges:2013eia,Berges:2013fga}, in the weak coupling regime of high occupancies $f >>1$, that the ratio of longitudinal to transverse pressures $P_L/P_T$ decreases with a power law in time-- albeit, this decrease is considerably slower than what one would expect from free streaming. Further, it appears that this  scaling behavior has a universal power law dependence in time, independent of whether the initial momentum distribution is prolate or oblate and independent of variations in the initial occupancy. Examining the single particle distributions extracted from the numerical simulations, we observe that they approach an attractor solution 
\begin{equation}
f(p_Z,p_T,\tau) = (Q_S\tau)^\alpha\, f_S\left( (Q_S\tau)^\beta p_T, (Q_S\tau)^\gamma p_Z\right) .
\end{equation}
Here we see that the distributions at different times can be scaled into a time-independent stationary function $f_S$ whose time dependence is only implicit through a rescaling of the transverse and longitudinal momentum scales with characteristic coefficients $\beta$ and $\gamma$. The system cools with an overall power of time characterized by $\alpha$. We find that our numerical simulations pick the values of 
$\alpha = -2/3$, $\beta=0$ and $\gamma = 1/3$ to good accuracy. The quoted values for the scaling exponents are precisely those predicted in the ``bottom up" thermalization scenario-henceforth BMSS.  

These results were unexpected because the BMSS scenario does not contain the effects of plasma instabilities which must be present in a weak coupling kinetic theory framework~\cite{Kurkela:2011ub,Berges:2015ixa,Mrowczynski:2016etf}. The numerical simulations indicate that an overpopulation of gluons in the infrared must suppress these late time plasma instabilities. The gauge theory results are corroborated by simulations of a weakly coupled but strongly self-interacting scalar theory with the same geometry which, remarkably, displays the same attractor solution~\cite{Berges:2014bba}. While the scalar theory does not know about plasma instabilities, it demonstrates a significant overpopulation in the infrared. In fact, it can be demonstrated clearly that this overpopulation in the scalar case leads to a non-equilibrium Bose-Einstein Condensate~\cite{Berges:2015ixa,Orioli:2015dxa}.

A numerical implementation of the BMSS bottom up kinetic scenario was shown recently to match smoothly to second order hydrodynamic at times on the order of a Fermi when the results are extrapolated to realistic couplings at RHIC \& LHC energies~\cite{Kurkela:2015qoa,Keegan:2016cpi}. We now therefore have a proof of principle realization of a heavy -ion collision all the way from its earliest instants to viscous hydrodynamics. However, caveat emptor! There are still several technical and conceptual hurdles to cross before this description can be declared fully robust. 

One test of this weak coupling scenario is to seriously address its implications for a variety of final states in heavy-ion collisions. An immediate application is to photon production; as an electromagnetic probe, it is sensitive to production mechanisms in the various stages of a heavy-ion collision. A state-of-the art computation of photon production employs weak coupling estimates of photon production from the QGP~\cite{Paquet:2015lta,Paquet}. Our prior discussion suggests that, for self-consistency, it is important to estimate this thermal weak coupling result relative to those from various stages of the Glasma. The pre-equilibrium contributions to photon production may be especially important in smaller systems~\cite{Shen:2016zpp}. Such studies should be extended to other final states (Onium production being an outstanding example) which may be sensitive to early time dynamics.

\subsection{The Glasma and the ridge}

The discovery of ridges in high multiplicity proton-proton collisions is now more than five years old. The outstanding question whether they are due to initial state or final state effects remains~\cite{Dusling:2015gta}. Experiments have demonstrated conclusively that, at low $p_T$, the dynamics is collective. This would seem to point to hydro as the mechanism; however, this conclusion is premature. For $p_T < Q_S$, as articulated, one is in the dense-dense kinematics, even for $p+p$ collisions. Computations of two particle correlations from Yang-Mills dynamics~\cite{Lappi:2009xa,Kovchegov:2012nd,Schenke:2015aqa,Chirilli:2015tea}, show many of the characteristic patterns at high multiplicities, such as, a) the mass ordering of $v_n$ coefficients, that are often attributed to flow~\cite{Schenke:2016lrs}, and b) the energy independence of ridge yields at a given $N_\text{charge}$~\cite{Dusling:2015rja}. At high $p_T > Q_S$, the Yang-Mills framework smoothly evolves to the Glasma graph picture, which describes the systematics of two particle correlations in the high $p_T$  regime quite well~\cite{Dumitru:2010iy,Dusling:2012iga,Dusling:2013qoz}.

The remaining outstanding questions are: i) whether Yang-Mills dynamics describes the collectivity seen at low $p_T$ in p+p and p+A collisions?  While this may be understood qualitatively~\cite{Kovner:2012jm,Dumitru:2014yza,Lappi:2015vta}, quantitative studies are computationally intensive and inconclusive thus far. ii) Whether the systematic trends seen in $p+A$,  $d+Au$ and $3He+Au$ collisions at RHIC can be understood. While the $v_2$ coefficients show an increasing pattern with system size, and $v_3$ is significant in 3He+Au collisions, as might be anticipated in a geometric scenario, it must be kept in mind that a $0-10$\% centrality selection corresponds to very different event activity ($N_\text{charge}$) for the three systems. In our initial state scenario, the $v_n$'s also grow with increasing event activity. iii) What are the characteristics of mini-jets in high multiplicity $p+p$ and $p+A$ collisions? While jets are rare, mini-jets are copiously produced, and form a significant fraction of the total multiplicity~\cite{Dusling:2012cg}. Therefore, subtracting their contribution (using minimum bias events) by  assuming the mini-jets are unmodified in high multiplicity events, and attributing the remainder to flow, appears to be contradictory. Why should some significant fraction of the multiplicity not be modified by flow? Such questions are relevant not just for the final state models but for initial state scenarios as well. 
 Understanding these systematics in detail, as well as long range rapidity correlations of other semi-hard final states such as open charm pairs, open charm-hadron pairs and photon-hadron pairs, will provide further insight into this fascinating topic. 

\section*{Acknowledgments}

I would like to thank the organizers of the Hard Probes conference for their kind invitation and
hospitality in Wuhan. This work was supported under DOE Contract No. DE-SC0012704.




\nocite{*}
\bibliographystyle{elsarticle-num}
\bibliography{Venugopalan_R}







\end{document}